\documentclass[
    ,final            
  ]
  {aipproc}

\layoutstyle{8x11single}

\usepackage{graphicx}

\begin{document}

\title{Recent Developments in Simulations of an Inverse Cyclotron for Intense Muon Beams}


\author{Kevin Paul}{
  address={Tech-X Corporation, Boulder, CO 80303, USA}
}

\author{Estelle Cormier-Michel}{
  address={Tech-X Corporation, Boulder, CO 80303, USA}
}

\author{Terrence Hart}{
  address={University of Mississippi-Oxford, University, MS 38677, USA}
}

\author{Donald Summers}{
  address={University of Mississippi-Oxford, University, MS 38677, USA}
}

\begin{abstract}
A number of recent developments have led to simulations of an inverse cyclotron for cooling intense muon beams for neutrino factories and muon colliders. Such a device could potentially act as a novel beam cooling mechanism for muons, and it would be significantly smaller and cheaper than other cooling channel designs. Realistic designs are still being explored, but the first simulations of particle tracking in the inverse cyclotron, with accumulation in the cyclotron core, have been done with electrostatic simulations in the particle-in-cell code VORPAL. We present an overview of the muon inverse cyclotron concept and recent simulation results.
\end{abstract}

\maketitle


\section{Introduction}

Muon beam cooling is a necessity in modern muon collider and neutrino factory designs, and the only known means of cooling muon beams quickly enough before their eventual decay is through the technique of \textit{ionization cooling}~\cite{Skrinsky:1981ht, Neuffer:1983jr, Neuffer:1994yv, Neuffer:1996eg, Palmer:1994km}.  In ionization cooling channels, muon beams lose energy while passing through a low-$Z$ material, such as liquid or gaseous hydrogen or lithium hydride.  Combined with transverse focusing and RF re-acceleration, this process quickly reduces the transverse emittance of the beam.  Emittance exchange techniques can be used to cool longitudinally as well.

A variety of muon cooling channels has been devised~\cite{Gallardo:1996a, Ankenbrandt:1999a, Alsharoa:2002wu, Palmer:2005a, Palmer:2007b}.  All of these channels are long and expensive, due mainly to the perceived need to keep the muon beam at its produced kinetic energy, around 100 MeV.  This keeps the beam at energies high enough to minimize the rate of energy loss in the material and also prevents the longitudinal emittance growth due to the steeply negative stopping power curve for energies below 100 MeV.  This requires constant alternation between low-$Z$ material energy loss and RF re-acceleration.  Such channels tend to be hundreds of meters long, resulting in decay losses of 10\% to 20\%, roughly equivalent to the same decay losses expected after 200 ns to 600 ns for a stopped muon beam.

In an inverse cyclotron, muons are injected into the cyclotron at the outer radius and proceed to spiral into the core of the cyclotron, losing energy in a low-$Z$, low-density gas such as hydrogen or a low-$Z$ foil in the central plane of the cyclotron.  After accumulation in the core, the muons are ejected out of the core in the direction perpendicular to the cyclotron plane and re-accelerated.  Such devices date back thirty years \cite{Simons:1988, Eades:1989, Eades:1989yd, Aschenauer:1992, DeCecco:1997, Summers:2005ij, Summers:2005vk, Bollen:2008a} and have been used for low-intensity muon and anti-proton beams.  

Inverse cyclotrons are much smaller and less expensive than currently proposed cooling channels, but the concept has not yet been proven viable for intense muon beams.  The outstanding issues involve whether the device can accept the initial, hot muon beam with minimal losses, and whether ionization rates in a gas or capture losses in a foil will be excessive.  Also, it is uncertain how significant the space charge forces will be as the muons coalesce in the core of the inverse cyclotron or during ejection from the core.  In this study, we address the effects of space-charge forces during accumulation of the beam in the core, where the muons are effectively stopped.

In the next section, we describe the simulation tools we use for this study, G4beamline and VORPAL.  In the following section, we describe simulations performed and results.  The last section entails our conclusions and plans for future work.

\section{Simulation Tools}

The simulations of the inverse cyclotron for this study were conducted with two very different codes: G4beamline and VORPAL.  G4beamline was used to explore and scan design parameters of the inverse cyclotron via single-particle tracking.  With G4beamline simulations, we were able to assess the rough acceptance of the device with much less effort than with VORPAL.  These parameters were then used for simulations with VORPAL, where the space-charge effects of the beam can be modeled by numerically solving the Poisson equation.

G4beamline~\cite{G4beamline, Roberts:2008a} is a particle tracking code developed by Tom Roberts at Muons, Inc.,~\cite{MuonsInc} built on top of the GEANT4~\cite{Agostinelli:2002hh} toolkit, written in C++.  It is capable of fully 3-dimensional charged-particle tracking in electric and magnetic fields, with the GEANT4 capabilities of tracking charged particles through matter, particle decay, and even particle production from high-energy impact with targets.  It has become a widely used tool within the muon beams community and is actively being developed for use in simulating the helical cooling channels being developed by Muons, Inc \cite{Yonehara:2006zs}.  

VORPAL is a fully-electromagnetic/electrostatic particle-in-cell (PIC) code, developed by Tech-X Corporation for modeling fully self-consistent electromagnetic laser-plasma interactions~\cite{1007846}.  Since its original inception, its capabilities have been significantly expanded to allow for purely electrostatic calculations~\cite{1007846b}, as well as many particle interactions, including field and impact ionization, particle-particle collisions, and fluid modeling.  VORPAL is an object-oriented C++ simulation framework for 1-D, 2-D, and 3-D parallel simulations of fully relativistic charged fluids and particles on a structured grid.  VORPAL can also combine fluid and kinetic modeling to provide hybrid simulation capability.

\section{G4beamline \& VORPAL Simulations}

\begin{figure}
  \includegraphics[height=.35\textheight]{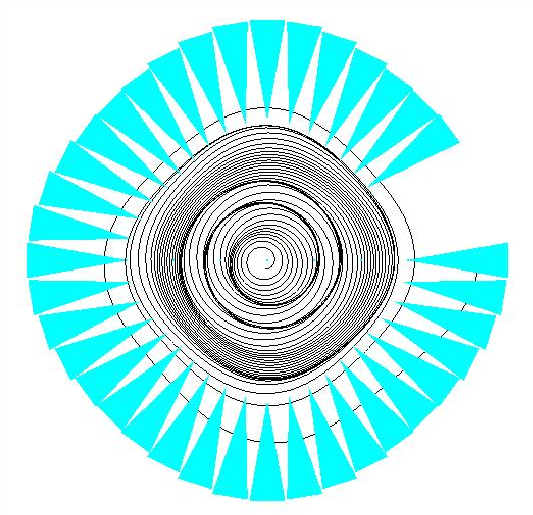}
  \caption{This figure shows the results of initial simulations with G4beamline of energy loss injection.  The outermost region of the cyclotron contains lithium hydride wedges, causing the beam to lose energy significantly faster in this region than in the interior of the cyclotron itself.  As a result, the beam loses enough energy in a single turn to have its orbital radius reduced by more than the diameter of the injected beam.  The black track in the figure shows the design orbit.}
  \label{fig:injection}
\end{figure}

The nominal design of the inverse cyclotron for muon cooling consists of three stages: (1) energy loss injection, (2) accumulation into the core, and (3) ejection.  In the first stage, the beam is injected into the cyclotron at the outer radius where thin wedges of solid low-$Z$ material, such as lithium hydride, are placed periodically around the ring.  The beam loses enough energy in the wedges such that its orbital radius in the cyclotron after a single turn is less than the injection radius by at least the beam diameter.  Figure~\ref{fig:injection} depicts initial simulations of this stage with G4beamline.

Once the beam has been injected into the cyclotron, the beam enters the second stage, spiraling into the core of the cyclotron as it loses energy in the low-density gas or foil.  A particle trap is placed at the core to contain the stopped beam before space-charge forces blow up the trapped bunch.  After the beam has had enough time to accumulate in the core---a small fraction of the muon lifetime---the electric trap fields are relaxed and an electrostatic kicker is applied to eject the muons from the core into a low-energy acceleration stage.  A full simulation of the inverse cyclotron requires simulation of all three of these stages.  In this article, however, we consider only accumulation during the second stage.

\begin{figure}
  \includegraphics[height=.5\textheight]{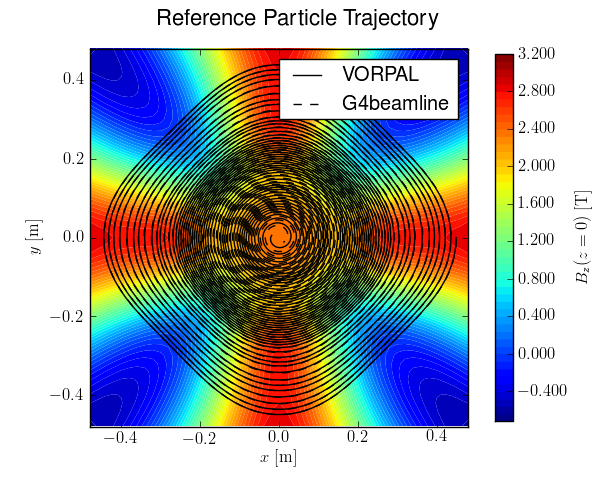}
  \caption{This figure shows the reference trajectory for the second stage of the inverse cyclotron, where the muons spiral into the core of the cyclotron.  One can see that the reference trajectory predicted by the two simulations codes, VORPAL and G4beamline, are slightly different near the core.  This is due to the slightly different energy loss rates predicted by the two codes, which differ most significantly at low energies.  These two orbits are superimposed on top of a 2D map of the magnetic field strength in the central plane of the cyclotron.}
  \label{fig:xyTraj}
\end{figure}

The magnetostatic field in the central plane of the cyclotron is assumed to take the following form,
\begin{equation}
B_z(r,\theta,z=0) = B_0 \exp\left(-\frac{r^2}{2\sigma^2}\right) + \frac{1}{2}\left(\frac{r}{a}\right)^{0.6} \left(1+\tanh\left(\frac{r-r_0}{\ell}\right)\right) \left(1 + f \cos(4\theta)\right) \;\; ,
\end{equation}
where the first term roughly describes a magnetic bottle in the core of the cyclotron and the second term describes the 4-sector cyclotron fields~\cite{ChaoTigner:1999}.  The coefficients of the in-plane field expressed above are $B_0 = 2.4$ T,  $\sigma = 0.2$ m, $a = 0.386$ m, $\ell = \frac{1}{15}$ m, $r_0 = 0.2$ m, and $f = \sqrt{2}$.  In the simulations, the above in-plane field is expanded to fourth order in $z$ to approximate the full field~\cite{Hart:2010a}.  Figure~\ref{fig:xyTraj}
shows the trajectory of the design particle in the inverse cyclotron fields described above, as predicted by the two simulations codes G4beamline and VORPAL.  The two trajectories differ slightly because the predicted energy loss in the two codes is slightly different.  This difference is considered small for this study, as the scattering and space-charge forces result in considerably larger effects.  For this simulation study, the interior of the cyclotron is assumed to be filled with $2.5 \;\rm{kg}/\rm{m}^3$ hydrogen gas.

Single-particle G4beamline simulations were then performed to find the design orbit for a muon with an initial momentum of 180 MeV/c.  This orbit was determined to start with a radius of 0.45 m at the center of one of the four sector bending magnets\footnote{This is merely for convenience since the energy-loss injection is not being modeled.}, chosen to lie on the $x$-axis as shown in Figure~\ref{fig:xyTraj}.  

Using G4beamline, single-particle scans were made over $(x, p_x, z, p_z, p)$ phase-space by varying only one of these parameters at a time.  The muons that make it into the 5 cm radius core were recorded, and their initial positions in phase-space are assumed to map out a region of acceptance.  We extract the widths of this region in each of the five dimensions of the scanned phase-space.  The radial width is found to be roughly 5 cm.  The transverse momentum width is found to be roughly $\pm20$ MeV/c, and the longitudinal width is found to be roughly $\pm20$ MeV/c.

Assuming a gaussian acceptance with these widths and a uniformly-distributed (i.e., ``beer-can'') beam with these same widths, we estimate $(\rm{erf}(1/\sqrt{2}))^5 = 15\%$ of the $2\times10^{12}$ initial muons ($3\times10^{11}$ muons) should make it to the core.  Initial VORPAL simulations find roughly $3.6\times10^{11}$ muons arrive in the core with such a beam, which is slightly better than our gaussian-based estimates but still quite small.  We, therefore, assume a uniformly-distributed beam of $2\times10^{12}$ muons with half of the above acceptance widths.  The beam is then assumed to be 100 ns is length, which is comparable to the length to a typical muon beam after production in a pion decay channel.  Table~\ref{tab:VORPALbeam} summarizes the beam properties used in the VORPAL simulations.
\begin{table}
\begin{tabular}{ccccccc}
\hline
      \tablehead{1}{c}{b}{Initial\\Radius\tablenote{In center of bending magnet}}
  & \tablehead{1}{c}{b}{Initial\\Momentum}
  & \tablehead{1}{c}{b}{Beam\\Radius}
  & \tablehead{1}{c}{b}{Beam\\Length}
  & \tablehead{1}{c}{b}{Transverse\\Momentum\\Spread}
  & \tablehead{1}{c}{b}{Longitudinal\\Momentum\\Spread}
  & \tablehead{1}{c}{b}{Initial\\Normalized\\Transverse\\Emittance} \\
\hline
0.45 m & 180 MeV/c & 2.5 cm & 100 ns & $\pm10$ MeV/c & $\pm10$ MeV/c & 0.15 mm-rad\\
\hline
\end{tabular}
\caption{This table summarizes the beam parameters used in the VORPAL simulations.}
\label{tab:VORPALbeam}
\end{table}
The transverse emittance of the initial beam is 0.15 mm-rad, or roughly 10\% of the initial transverse emittance expected from typical pion decay channels.  Based on the same gaussian assumptions described above, such a beam should result in 79\% of the muons in the core ($1.58\times10^{12}$ muons).  It should be noted that, while this simulated beam is much smaller than typical muons beams after their production, pre-cooling can be done to reduce the emittance such that the actual muon beam meets these acceptance criteria.  However, we are also exploring modified cyclotron designs to increase the acceptance of the device so that less, or possibly no, pre-cooling is necessary.

Using these beam parameters, VORPAL was used to conduct simulations of the inverse cyclotron with and without scattering (resulting in beam heating) and space-charge forces.  To prevent the inevitable blow-up of the bunch in the core when space-charge forces are enabled, an approximation to a simple electrostatic trap was added to the simulation by adding a point-charge-like field\footnote{To prevent excessively large kicks to the muons near the center, we cut off the growth of the trap field within a radius of 2 cm.} equal to the field produced from $2\times10^{12}$ muons at the center of the cyclotron.  We run the simulation for 400 ns, which accounts for the roughly 300 ns spiral time of the design particle and the 100 ns length of the beam.
\begin{figure}
  \includegraphics[height=.4\textheight]{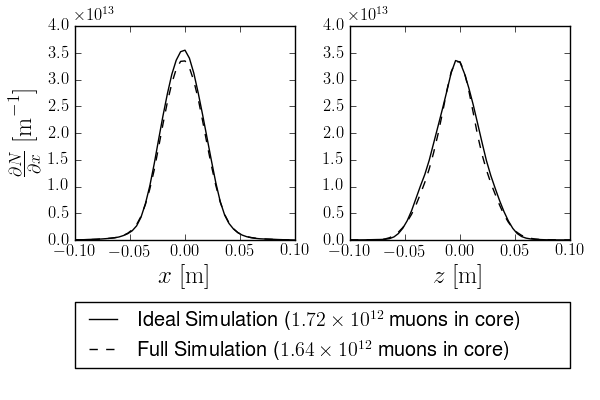}
  \caption{This figure shows the distributions of the accumulated muons in the core after 400 ns for various VORPAL simulation configurations.  The \textit{Ideal} simulation denotes the full beam VORPAL simulation described in the text with scattering disabled, space-charge forces disabled, and no electrostatic trap in the core.  The \textit{Full} simulation denotes the full beam VORPAL simulation described in the text with space-charge forces and scattering enabled and an approximated electrostatic trap in the core.}
  \label{fig:coreHists}
\end{figure}

Figure~\ref{fig:coreHists} shows the results of two different configurations, one with scattering and space-charge forces disabled (\textit{Ideal}) and one with scattering and space-charge forces enabled with an approximated electrostatic trap in the core (\textit{Full}).  The \textit{Ideal} simulation shows $1.72 \times 10^{12}$ muons accumulating in the core, of an initial $2\times10^{12}$ initially simulated.  This means roughly 14\% of the muons are lost due to the magnetic field configuration alone.  When space-charge forces and scattering are enabled, and the approximated electrostatic trap is added to the simulation (\textit{Full}), an additional 4\% (18\% total) of the muons are lost.  This suggests that the scattering effects are not terribly adverse, and that the space-charge forces can be effectively compensated with an electrostatic trap (assuming the trap field strengths are not prohibitively large to construct and operate).  It also suggests that space-charge forces do not have a significant effect outside of the cyclotron core.

\section{Conclusions \& Future Work}

The simulations performed in this study are the first of their kind for the muon inverse cyclotron.  They represent the first full beam simulations of accumulation in the core of the cyclotron with space-charge forces enabled.  The results are very encouraging.  

Based on parameter scans with G4beamline, we were able to develop preliminary designs for the inverse cyclotron that performed well in full beam simulations with VORPAL.  The VORPAL simulations suggest that space-charge forces are not a significant problem outside the core, and these simulations confirm the G4beamline-tested design.  A 0.15 mm-rad transverse emittance beam of $2\times10^{12}$ muons was successfully accumulated in the core of the cyclotron with only 18\% loss after 400 ns.  The expected decay losses in this time are roughly 16\%, resulting in a total of 34\% losses.  While the initial beam used for these simulations has a much smaller emittance than initial muons beams in typical neutrino factory and muon collider designs, pre-cooling can be applied to the beam to reduce the emittance until it meets the acceptance criteria.

Future work will require full end-to-end simulations of the inverse cyclotron, including complete simulations of all three stages.  With these full simulations, following ejection and initial re-acceleration of the beam, we will be able to make the first estimates of the cooling capability of the muon inverse cyclotron.  Designs of the cyclotron fields continue to be explored with G4beamline in the hopes of finding a design with a larger acceptance, such that less (or no) pre-cooling is necessary before injection into the inverse cyclotron.  Realistic designs of the electrostatic trap and ejection from the core have been simulated in VORPAL, but these trap designs have yet to be incorporated into the VORPAL simulations performed for this study.  Initial acceleration after ejection must also be simulated.  Additionally, G4beamline simulations are being performed to optimize the energy-loss injection scheme, but this scheme has not been implemented in the VORPAL simulations to date.  Estimates of losses due to muonium formation (for $\mu^+$) and muon atomic capture (for $\mu^-$), as well as estimates of ionization rates and the behavior of the resulting plasma, must also be computed to fully assess the inverse cyclotron concept.  However, we believe that full end-to-end simulations of the muon inverse cyclotron are feasible in the near future.


\begin{theacknowledgments}

We gratefully acknowledge the support from the Department of Energy, Office of High Energy Physics, SBIR grant DE-FG02-08ER85044, and the National Science Foundation Award 757938.  This work is also supported in part by Tech-X Corporation, Boulder, CO, and the University of Mississippi-Oxford, University, MS.

We would also like to acknowledge the Tech-X Corporation VORPAL development team:
D.~Alexander,
K.~Amyx,
T.~Austin,
G.~I.~Bell,
D.~L.~Bruhwiler,
R.~S.~Busby,
J.~Carlsson,
J.~R.~Cary,
E.~Cormier-Michel,
Y.~Choi,
B.~M.~Cowan,
D.~A.~Dimitrov,
M.~Durant,
A.~Hakim,
B.~Jamroz,
D.~P.~Karipides,
M.~Koch,
A.~Likhanskii,
M.~C.~Lin,
J.~Loverich,
S.~Mahalingam,
P.~Messmer,
P.~J.~Mullowney,
C.~Nieter,
K.~Paul,
I.~Pogorelov,
V.~Ranjbar,
C.~Roark,
B.~T.~Schwartz,
S.~W.~Sides,
D.~N.~Smithe,
A.~Sobol,
P.~H.~Stoltz,
S.~A.~Veitzer,
D.~J.~Wade-Stein,
G.~R.~Werner,
N.~Xiang,
C.~D.~Zhou.
\end{theacknowledgments}


\bibliographystyle{aipproc}   

\bibliography{bibliography}

\end{document}